\documentclass[conference]{IEEEtran}
\IEEEoverridecommandlockouts
\usepackage{amsmath,amssymb,amsfonts}
\usepackage{algorithmic}
\usepackage{graphicx}
\usepackage{textcomp}
\usepackage{xcolor}
\usepackage{natbib}
\usepackage{booktabs}
\usepackage{adjustbox}
\usepackage{url}
\usepackage{hyperref}
\usepackage{pifont} 
\def\BibTeX{{\rm B\kern-.05em{\sc i\kern-.025em b}\kern-.08em
    T\kern-.1667em\lower.7ex\hbox{E}\kern-.125emX}}
\begin{document}

\title{ClassVision: AI-Powered Classroom \\Attendance System
}

\author{\IEEEauthorblockN{Ankit Kumar Aggarwal}
\IEEEauthorblockA{\textit{Artificial Intelligence} \\
\textit{Yeshiva University}\\
NY, USA \\
aaggarwa@mail.yu.edu}
\and
\IEEEauthorblockN{Veerabhadra Rao Marellapudi}
\IEEEauthorblockA{\textit{Artificial Intelligence} \\
\textit{Yeshiva University}\\
NY, USA \\
vmarella@mail.yu.edu}
\and
\IEEEauthorblockN{Ovadia Sutton}
\IEEEauthorblockA{\textit{Artificial Intelligence} \\
\textit{Yeshiva University}\\
NY, USA \\
osutton@mail.yu.edu}
\and
\IEEEauthorblockN{Youshan Zhang}
\IEEEauthorblockA{\textit{Artificial Intelligence} \\
\textit{Yeshiva University}\\
NY, USA \\
youshan.zhang@yu.edu}
}


\maketitle

\begin{abstract}
Students and working professionals have to go through the attendance process every day. Traditional methods of marking attendance using pen and paper or online platforms are human-intensive and time-consuming. To address the challenges in manual attendance processes, this research explores the use of face detection (FD) and face recognition (FR) technology to automate the attendance process, particularly in educational settings, and build a ClassVision course attendance system. We also propose an automated attendance system featuring a human-computer interaction (HCI) and user-friendly web interface that utilizes real-time image processing to identify and recognize students in classrooms and automatically record their attendance. We identified RetinaFace as the best face detection model, and when combined with Face Recognition for verification, it provided the most promising results with a cropped embedding of 50x50 pixels. Our proposed framework achieves a detection accuracy of 99.4\% and a recognition accuracy of 90.3\%. Unlike deep learning methods that require students to be present closely in front of the camera to capture attendance, our pioneering research aims to capture attendance seamlessly during ongoing classes, redefining the framework of attendance tracking systems. Source code is available at~\url{https://tinyurl.com/Course-Attendance-Robot}.

\end{abstract}



\section{Introduction}
The automation of taking attendance is important for education, corporate, and government institutions to track the performance of students and employees. 
Traditional pen and paper methods are time-consuming and prone to errors and they can enable fraudulent practices like impersonation~\cite{aggarwal2020principal}. In educational institutes, the conventional approach, calling out names or passing around a pen and paper disrupts lectures and wastes time. Reliance on attendance lists leads to frustration for both students and instructors~\cite{xu2019application}. Furthermore, paper-based systems pose security risks, as documents can be misplaced or pilfered, complicating the task of retrieving attendance records. Addressing these challenges necessitates the adoption of alternative methods that ensure secure and efficient attendance management without human intervention. Biometric systems, leveraging technologies like fingerprint recognition, represent a promising avenue for automating attendance processes.

Modern attendance systems utilize biometric data, specifically facial features, to precisely identify individuals and log their attendance. These systems offer improved accuracy and efficiency compared to conventional methods such as sign-in sheets or roll calls~\cite{omari2020attendance}. However, there are multiple issues in automating attendance tracking in classroom settings. It involves research on relevant camera devices with capabilities to deploy the best model for capturing classroom images during lectures. It will be used to capture random images during lectures and employ facial recognition algorithms to identify and record their attendance. This system prioritizes user-friendliness and efficiency, utilizing readily available devices that can be deployed to capture attendance. By continually analyzing facial data, this approach aims to improve the accuracy and effectiveness of attendance-tracking methods~\cite{anitha2020face}.

\begin{figure}[ht]
  \centering
  \includegraphics[width=\linewidth, height=5cm]{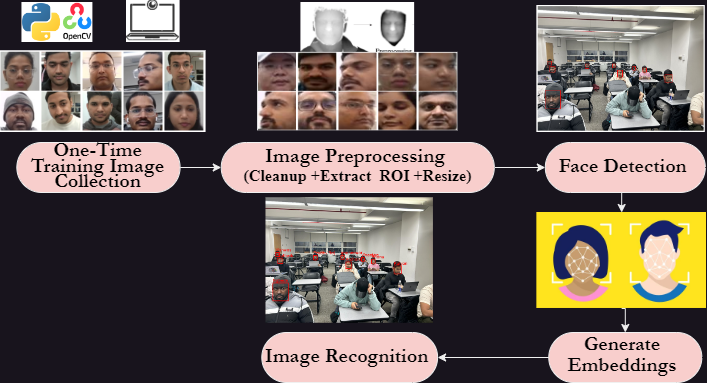}
  \caption{ClassVision Framework.}
  \label{fig:ClassVision Framework}
\end{figure}

Facial recognition technology raised privacy concerns over data collection and potential algorithm bias~\cite{gornale2020classroom}. We followed robust data protection measures to safeguard privacy and prevent unauthorized access. Hence we could only collect test classroom images from the volunteers from specific rooms in this research. This study introduces an attendance tracking system using facial recognition technology, which can be implemented using cost-effective high-resolution cameras. It utilizes a five-stage face recognition ClassVision framework, as outlined in Fig.~\ref{fig:ClassVision Framework} for achieving remarkable results on group face recognition in classroom settings.

The first step involves a one-time training image collection of dimensions 112x112. In the second step, training images will be processed to extract the region of interest (ROI), which will then be used to create known face embeddings. In the third step, test images of the classroom will be processed by the retina face to detect all the faces in the picture. In the fourth step, embedding will be generated using face recognition. Lastly, the comparison will be carried out between the facial embeddings from test images and known face embeddings from training images. We have also created a user-friendly interface, empowering administrators and verified users to manage and derive insights from attendance records efficiently. Our research is primarily based on images captured within a classroom environment, where students in the front demonstrate more facial features compared to those sitting at the back. Our contributions are threefold.


\begin{itemize}
    \item We are the first to develop a ClassVision framework and design a human-computer interaction interface to maximize the possibility of automatic course attendance in classroom settings.  

    \item We can capture face images from various angles (front, left, right, up, and down) to form a basis for training images, for which we will retrieve students' faces and names to take attendance with classroom images.

    \item We propose to extract the region of interest (ROI) from training images and identify the optimal ROI size to be 50x50 pixels. We then employ RetinaFac as the face detection model and python's face recognition method to achieve state-of-the-art attendance accuracy.
\end{itemize}
 

\section{Related Work}

In 1966, Bledsoe et al. pioneered facial recognition technology and highlighted the challenges faced during initial attempts. Emphasis was placed on using digital images or video to match individuals against a facial database~\cite{kumar2022face}. This technology has continuously evolved with improved algorithms for faster and more precise recognition.

\subsection{Face Detection}
Biometric systems, utilizing fingerprint and iris recognition technology, offerd enhanced security but are expensive and required more time for attendance capture. Saad et al.~\cite{FD31} considered multiple bands in the visible and near-infrared range, and explored their potential compared to monochromatic events and conventional multispectral imaging for face detection. Yuntao et al.~\cite{FD32} proposed LighterFace, in which two pre-trained convolutional neural networks were combined, namely Cross Stage Partial Network (CSPNet), and ShuffleNetv2. Additionally, Global Attention Mechanism (GAMAttention) was used to compensate for the accuracy loss. A facial spoofing detection approach based on weighted deep ensemble learning that combined the strengths of two powerful deep learning architectures, DenseNet201 and MiniVGG~\cite{FD34}. Another solution was proposed using an interpolation-based image diffusion augmented by transfer learning of a MobileNet convolutional neural network~\cite{FD35}. Jaffer et al.~\cite{FD36} improved YOLOv8n architecture and proposed two versions. With yolov8n-v1, integrating ResNet for better feature extraction and SPPCSP modules for efficiency. Yolov8n-v2 added Ghostconv and ResNet Downsampling. Face detection is crucial as it provides a reliable and secure method of identifying individuals, reducing the risk of fraud attendance. It also ensures that only authorized individuals are recorded, ensuring accuracy and accountability. 

\subsection{Face Recognition}

Advancements in deep learning techniques allowed researchers to introduce ArcFace, a novel loss function crafted specifically for face recognition tasks. This function projected facial features onto a hypersphere to improve distinguishability~\cite{gornale2020classroom}. 
Erfan et al.~\cite{FR21} modified the feature vectors of angular face images that involved segmenting the angular face image elements using a fine-tuned DeepLabv3 network. 
Charoqdouz et al.~\cite{FR23} proposed feature extraction from several angular faces using a deep learning-based fusion technique for face recognition. Fadi et al.~\cite{FR24} developed an extremely lightweight and accurate FR solution, namely PocketNet, a neural architecture search to develop a new family of lightweight face-specific architectures based on knowledge distillation (KD). John et al.~\cite{FR26} proposed a Neural Architecture Classifier (NAC) to avoid training architectures that would not have good performance, based on knowledge of previously trained architectures. A genetic algorithm (GA) was used for evolution, and a search space. Another team~\cite{FR27}, presented EdgeFace, which is a lightweight and efficient face recognition network inspired by the hybrid architecture of EdgeNeXt. 
By effectively combining the strengths of both CNN and Transformer models and a low-rank linear layer, EdgeFace achieved excellent face recognition performance optimized for edge devices~\cite{c6}.
Ghost modules utilized a series of inexpensive linear transformations to extract additional feature maps from a set of intrinsic features, allowing for a more comprehensive representation of the underlying information~\cite{c10}. Face recognition enhances the ability of attendance systems by providing the ability to match detected faces against known embeddings, enabling personalized attendance records and add additional security to the whole attendance records.

\subsection{Attendance Systems}
Traditional attendance recording methods were inefficient, prompting the exploration of automated systems. While Radio Frequency Identification-based (RFID) offered an initial solution, its limitations in accuracy, speed, and security led to a shift towards biometric attendance systems~\cite{mishra2022industry}. Student Attendance System integrated with facial recognition technology employing the Personal Component Analysis (PCA) algorithm~\cite{AS11}. Chang et al.~\cite{AS16} used an asymmetric face recognition technique which is crucial when names and faces lack a one-to-one correspondence. It introduces AFRM, which uses HOG, SVM, ConvFF, and KNN to label faces accurately. A real-time attendance system employing video-based face recognition was developed that focused on enhancing check-in accuracy, ensuring system reliability, and minimizing instances of truancy~\cite{ahmed2020comparative}. A CNN-based approach for monitoring student attendance in smart classrooms integrates IoT and edge computing for efficient data processing, surpassing conventional methods in face detection and recognition accuracy~\cite{farouk2022proposed}. A machine learning approach shows the  application in real-world scenarios like schools and traffic monitoring systems~\cite{ramasane2023enhanced}. Unlike traditional methods or some automated systems that require close proximity to the camera or suffer from slow processing speeds, our system leverages efficient algorithms for real-time processing. This enables seamless attendance-taking during ongoing classes without disrupting the flow. Furthermore, the framework is highly scalable and flexible, designed to accommodate various classroom sizes and adaptable to the number of students. Its user-friendly web interface sets it apart from less versatile solutions. By focusing on a region of interest with dimensions of 50x50 to create precise face embeddings, our system can accurately capture and correctly label the faces of students even in the back of the classroom, ensuring comprehensive coverage and reliability.

\section{Data Collection}

We collected a dataset from scratch. A good-quality dataset is one of the crucial aspects of our research, as these images will be used as the ground truth for evaluating the accuracy of the face recognition model. The dataset used in this project comprises images collected from multiple students. The whole data collection process is primarily divided into three phases.

\paragraph{Phase-1: Student Consent}
To ensure the ethical treatment of student data, we addressed privacy concerns upfront in our facial recognition system development. We obtained written consent through a No Objection Letter for the Collection of Student Images. This letter clearly explained the purpose of the research project – training and testing the face recognition algorithms – and participation was entirely voluntary, only after securing informed consent from a total of thirty-six (36) students enrolled in the NLP and AI classes who expressed willingness to participate in our research endeavor. 


\paragraph{Phase-2: Train Dataset}
For training our face recognition model, we utilized OpenCV (Open Source Computer Vision Library) to capture and preprocess images for face recognition. It consists of more than 2500 optimized algorithms~\cite{omari2020attendance}. These algorithms are capable of detecting and recognizing faces from an image. As OpenCV was designed for real-time application and image processing, we used it to capture photos in real-time through a webcam.

\begin{figure}[ht]
  \centering
  \includegraphics[width=\linewidth, height=5cm]{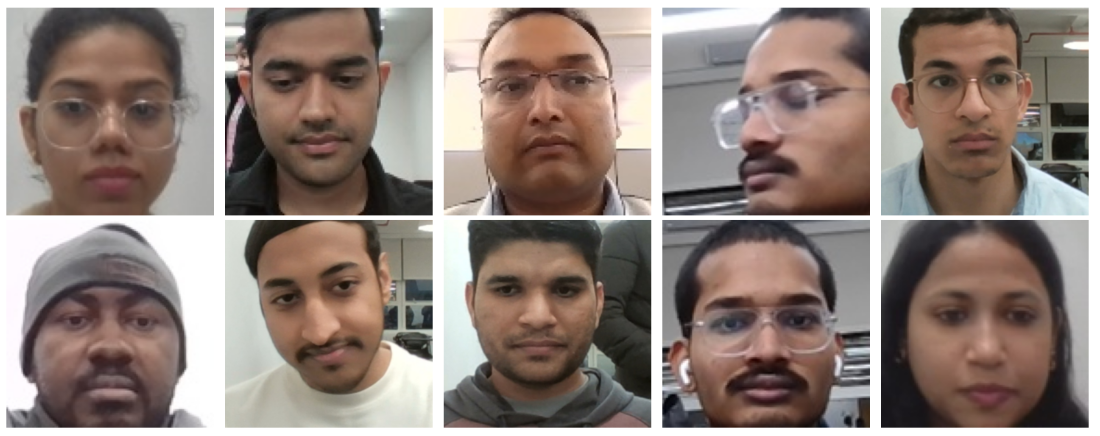}
  \caption{Training Images of Dimension 112x112.}
  \label{fig:training112}
\end{figure}

The Python script begins by loading a pre-trained Haar cascade classifier for face detection. Using this classifier, the code defines a function \verb|face_extractor| to detect faces in a given image and extract them for further processing. Upon initialization of the webcam, the user is prompted to input their name. The code then automatically creates a folder based on the provided name to store the captured images. During the image capture process, faces are continuously detected and extracted from the webcam feed. Each detected face is resized to match a resolution of 112x112 pixels and saved as a JPG file within the designated folder. The process continues until either 100 images are captured or the user terminates the capture process. Finally, the program releases the webcam and closes all windows, completing the image sampling process. The final training dataset encompassed 100 images for a sample size of 36 students, totaling 3,600 images. See Fig.~\ref{fig:training112} for samples.

\paragraph{Phase-3: Test Dataset}
The test dataset continued each week for up to 8 weeks by collecting 10 to 15 class group image samples under different conditions by exhibiting a variety of poses, backgrounds, lighting conditions, and facial expressions. By the end of 8 weeks, the final test dataset consisted of 69 class group photographs that were captured to evaluate the individual recognition of students' faces within group settings. Table~\ref{tab:dataset_summary} summarizes the dataset collected for training and testing images, including their respective dimension information. 

\begin{table}[htbp]
\centering
\caption{Dataset Summary.}
\resizebox{\columnwidth}{!}{%
\begin{tabular}{@{} *{7}{p{1cm}} @{}}
\toprule
Dataset & Image Count & Dimension & Mean Size (bytes) & Mean Width (bytes) & Mean Height (bytes) \\
\midrule
Training & 3600 & 112x112 & 13073 & 112 & 112 \\
Test & 69 & Mixed & 668148 & 1896 & 1422 \\
\bottomrule
\end{tabular}}
\label{tab:dataset_summary}
\end{table}

Alongside collecting images of students, we also measured the dimensions of various classrooms to assess the impact of classroom size on image accuracy. However, later, we decided to conduct our testing in the largest classroom and evaluate the results. Table~\ref{tab:classroom_dimensions} provides a summary of the different available classrooms and their respective dimensions.

\begin{table}[htbp]
\centering
\caption{Summary of Classroom Dimensions.}
\resizebox{\columnwidth}{!}{%
\begin{tabular}{@{} *{6}{p{1cm}} @{}}
\toprule
Building No. & Floor & Room No. & Height (ft) & Length (ft) & Width (ft) \\
\midrule
215 & 2 & 208 & 10 & 45 & 12 \\
215 & 3 & 302 & 10 & 45 & 12 \\
215 & 2 & 209 & 10 & 25 & 12 \\
245 & 1 & 102 & 10 & 20 & 15 \\
215 & 4 & 405 & 10 & 20 & 12 \\
Commas & G & - & 10 & 60 & 12 \\
\bottomrule
\end{tabular}}
\label{tab:classroom_dimensions}
\end{table}

During the data collection process, we also explored various camera options capable of supporting the deployment of our model to fully automate attendance management with limited human intervention. While investigating available camera options, our primary objective was to identify a camera that could be easily integrated with our code for image capturing, detection, and recognition. Later on, we decided to use laptop cameras for the first phase of research, and using specific high-resolution cameras will be explored in the next phase.

\section{Proposed Method}

In this section, we present a detailed description of the steps followed to develop the ClassVision solution to achieve state-of-the-art performance in facial recognition for a group of students, which is used to capture attendance without any manual intervention or disruption to the ongoing lecture. We formulate a unique combination and sequence of steps using various face detection and face recognition algorithms. After the successful collection of training data, the overall process is mainly divided into five main stages: 1) One-Time Training Image Collection 2) Extract Region of Interest (ROI) and resizing to 50x50, 3) Face Detection (Test Images), 4) Generate Embedding of detected faces, and 5) Image Recognition. The key to the process is the proper collection of training data followed by specified operations on training images to generate quality embeddings.  We first briefly introduce the Face Detection and Face Recognition models.

\subsection{Face Detection}

We applied RetinaFace~\cite{retinaface} as the best model for face detection, which is a robust single-stage face detector that performs pixel-wise face localization on various scales of faces by taking advantage of joint extra-supervised and self-supervised multi-task learning.

\subsection{Face Recognition}

We used a Face Recognition Python library that can be used to recognize and manipulate faces. It is built on top of Dlib~\cite{ageitgey_face_recognition}, a modern C++ toolkit that contains machine-learning algorithms and tools for creating complex software in C++ to solve real-world problems. As a part of facial recognition, after the facial images have been extracted, cropped, resized, and often converted to grayscale, the face recognition algorithm takes on the task of identifying features that most accurately represent the image. The face recognition systems can operate basically in two modes.
\begin{itemize}
\item Verification (1:1 Comparison): This method compares a captured facial image with a single reference image associated with a specific user seeking authentication. This process confirms the user's claimed identity.

\item Identification (1:N Comparison): This method scans a captured facial image against a larger dataset containing multiple user images. The objective is to identify the specific user from the dataset whose facial features match the captured image. This process establishes the user's identity without prior knowledge.
\end{itemize}

\subsection{Data Preparation}

During data preparation, we ensure that the training data is properly collected, with all images containing the faces of the respective students. We observed certain cases where some of the captured images contain more than one face, some of them have only a small portion of the face captured, and the rest of the image contains an empty background. We also encountered cases where some students who wear caps and also have thick beards create bias in the database, as they only return a small region of interest. There is a chance that these kinds of training images can easily confuse the model when dealing with a person having hair and thick beards on the face. Therefore, it's important to clean all such images and prepare a clean training dataset to avoid these kinds of biases during prediction. Once this manual cleanup is done, we will proceed to the next step, which is to extract the region of interest.

\subsection{Region of Interest (ROI)}

\begin{figure}[ht]
  \centering
  \includegraphics[width=\linewidth, height=5cm]{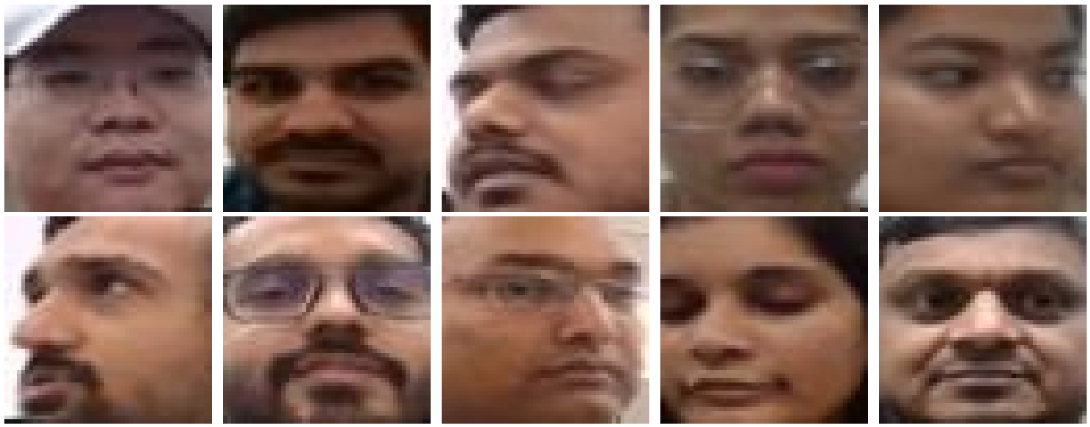}
  \caption{Region of Interest (ROI) of Dimension 50x50.}
  \label{fig:training50}
\end{figure}

After manually cleaning up the previous step, we automated the process of extracting the region of interest, performing resizing, and storing the images in a single directory with new names. We started by reading the training images, which are stored in a source folder containing images of students, with each student's images in a subfolder. We traversed each folder, processed each image, and copied it to the destination folder. During the copying process, we utilized Retinaface~\cite{retinaface} to identify the facial landmarks region in each image, extract the region of interest, resize it from 112x112 to 50x50, and save it to the destination folder. In the source folder, all images are stored with random numbers, while in the destination folder, we store each processed image with a name containing the folder name from which the file was copied (i.e., the student’s name), coupled with the original name of the image, (see Fig.~\ref{fig:training50}) for sample region of interest.

\subsection{Generate Embedding}

To generate embeddings, we defined a function called "load or generate embeddings." This function , denoted as  $f$, serves as an embedding extractor, allowing us to efficiently manage the embeddings and the corresponding names of the training images.  The introduction of this function aids in improving the overall execution time of predictions. Specifically, $f(x_{r}^{i})$ represents the embedding for the $i$th training image, while $f(x_{e}^{j})$ denotes the embedding for the $j$th test image. Here, $i$ represents one sample training image, and $j$ represents one sample test image. The function $f$ iterates through each of the student images available in the processed image folder and then loads them. Each loaded image detects the face by leveraging a face recognition library, which retrieves the facial embedding along with the corresponding image label. The label of the image contains the student's name, extracted from the file name, by considering only the part of the image until the first underscore. This process continues for each image in the folder. After processing all the images, the embeddings, along with the image names, are stored in a pickle file for future use. Generating the pickle file helps in performing quicker predictions if embeddings have already been generated for the training images. The process of generating embeddings for each image will be skipped if there already exists a pickle file with known embeddings. This function can easily be called to load or generate embeddings and store them in the variables `known face encodings' and `known face names'.

\subsection{Face Detection and Face Recognition}

It is the procedure of identifying a human face in an image and enclosing it in a box to isolate it from the rest of the image, followed by providing a name for the identified face. After successfully generating known face encodings and known face names, we will perform face detection and recognition on a set of classroom images, which were manually collected during the data collection stage. To achieve efficient face detection and recognition, we will first utilize the RetinaFace library to detect the facial landmarks of all the students in a classroom image. While performing face detection, each of the test images is loaded, and after successfully detecting facial locations, they are then converted to a format suitable for face recognition. This is followed by using a facial recognition library to extract the embeddings of detected faces, which will then be used to compare against the known face embeddings using the Euclidean distance metric to find the closest match, as shown in the equation: 


$EuclideanDistance(x_\text{detected}, x_\text{known}) = x_\text{detected} - x_\text{known}$

Once a successful match is found, a rectangular bounding box will be drawn around the face, and a label will be added to identify the face on top of the corresponding bounding box.

\subsection{Human Computer Interaction Interface}

After finalizing the best-proposed architecture, we delve into the process of creating an intuitive and user-friendly graphical user interface that can be seamlessly used by instructors to perform facial detection and recognition on the captured images. This interface provides the instructor with the facility to quickly load the image and visually verify any discrepancies in the captured attendance data. The portal provides various functionalities such as adding a course, viewing attendance for a specific day, and course selection to easily navigate and select images captured in a particular course. It also empowers instructors with the flexibility to manually review and ensure accuracy by addressing any discrepancies or overriding the automatically captured attendance status when necessary. Once the instructors are satisfied with the attendance record, they can easily download the attendance file in CSV format for further analysis and record-keeping. Here is a quick glimpse of the main screens from the user-friendly web portal, showing an intuitive application design flow.

\subsubsection{Add Course}

\begin{figure}[ht]
  \centering
  \includegraphics[width=\linewidth, height=5cm]{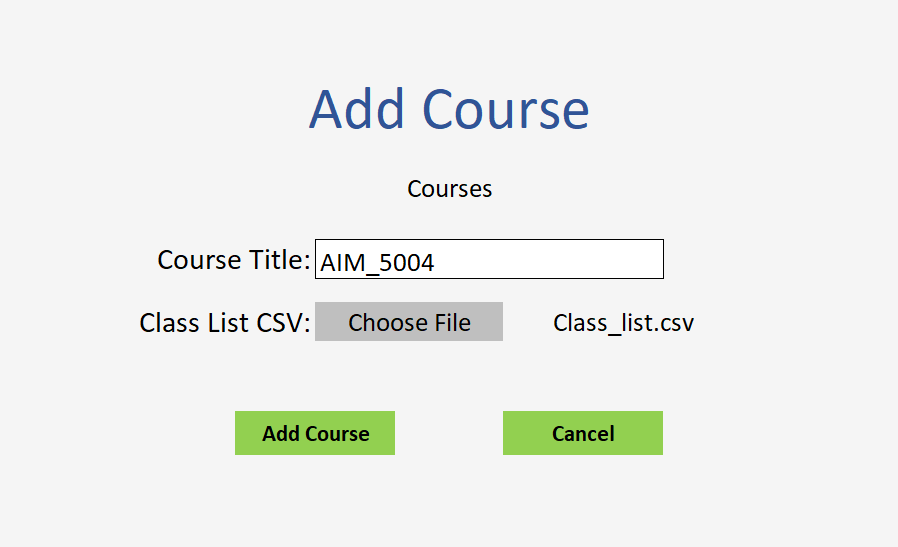}
  \caption{Screen of facilitate Course Addition.}
  \label{fig:AddCourse}
\end{figure}

The add course screen (see Fig.~\ref{fig:AddCourse}) can be used to add a new course to the list. It requires loading a CSV file that contains the names of the class students. 

\subsubsection{Capture Attendance}

\begin{figure}[ht]
  \centering
  \includegraphics[width=\linewidth, height=5cm]{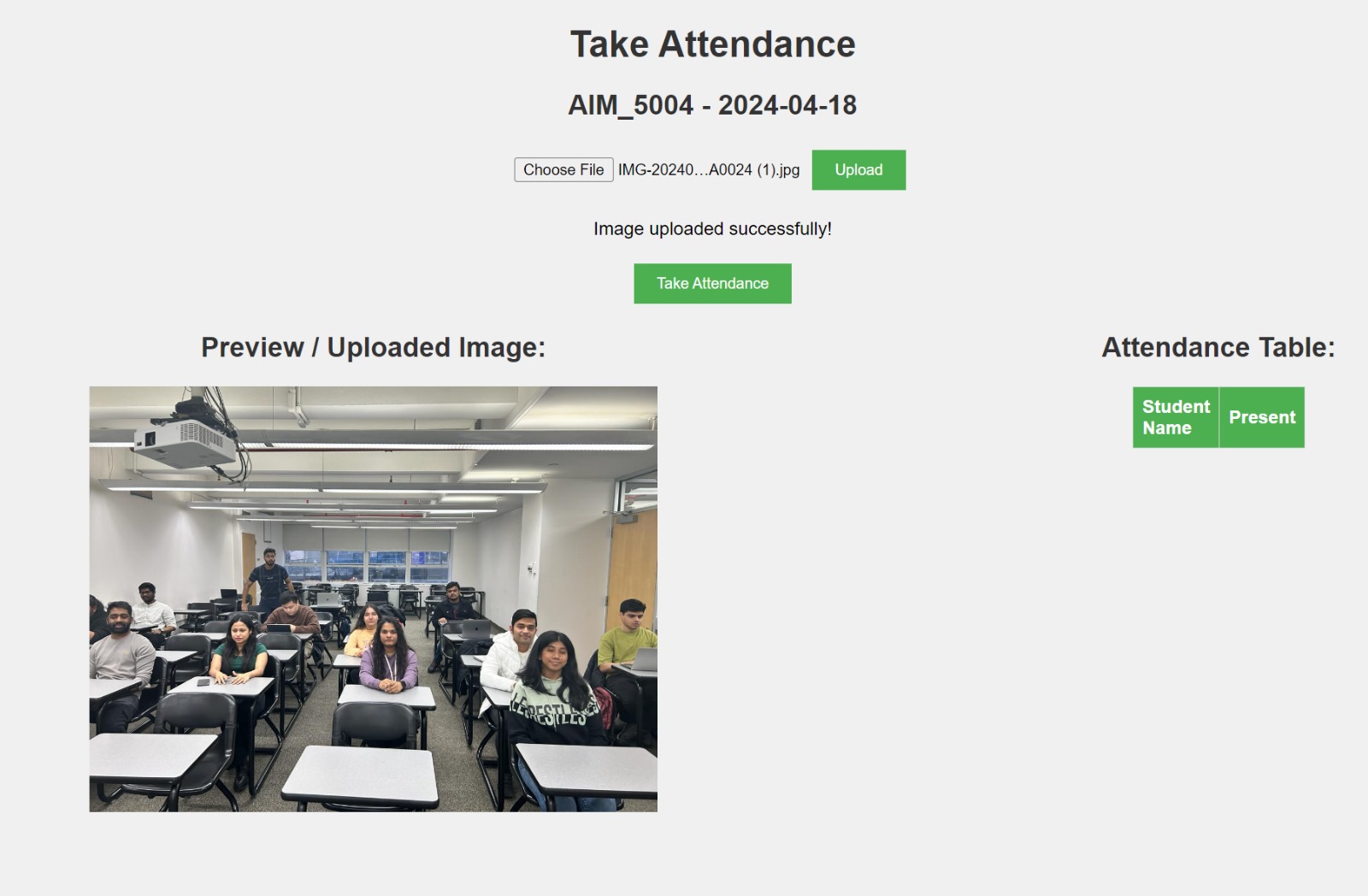}
  \caption{Screen of capturing attendance.}
  \label{fig:CaptureAttendance}
\end{figure}

The capture attendance screen (see Fig.~\ref{fig:CaptureAttendance}) provides users with the flexibility to select an image, perform face detection, and take attendance. It provides a visual overview of the detected faces and their corresponding named labels, which can easily be verified against the attendance table.

\subsubsection{Record Attendance}

\begin{figure}[ht]
  \centering
  \includegraphics[width=\linewidth, height=5cm]{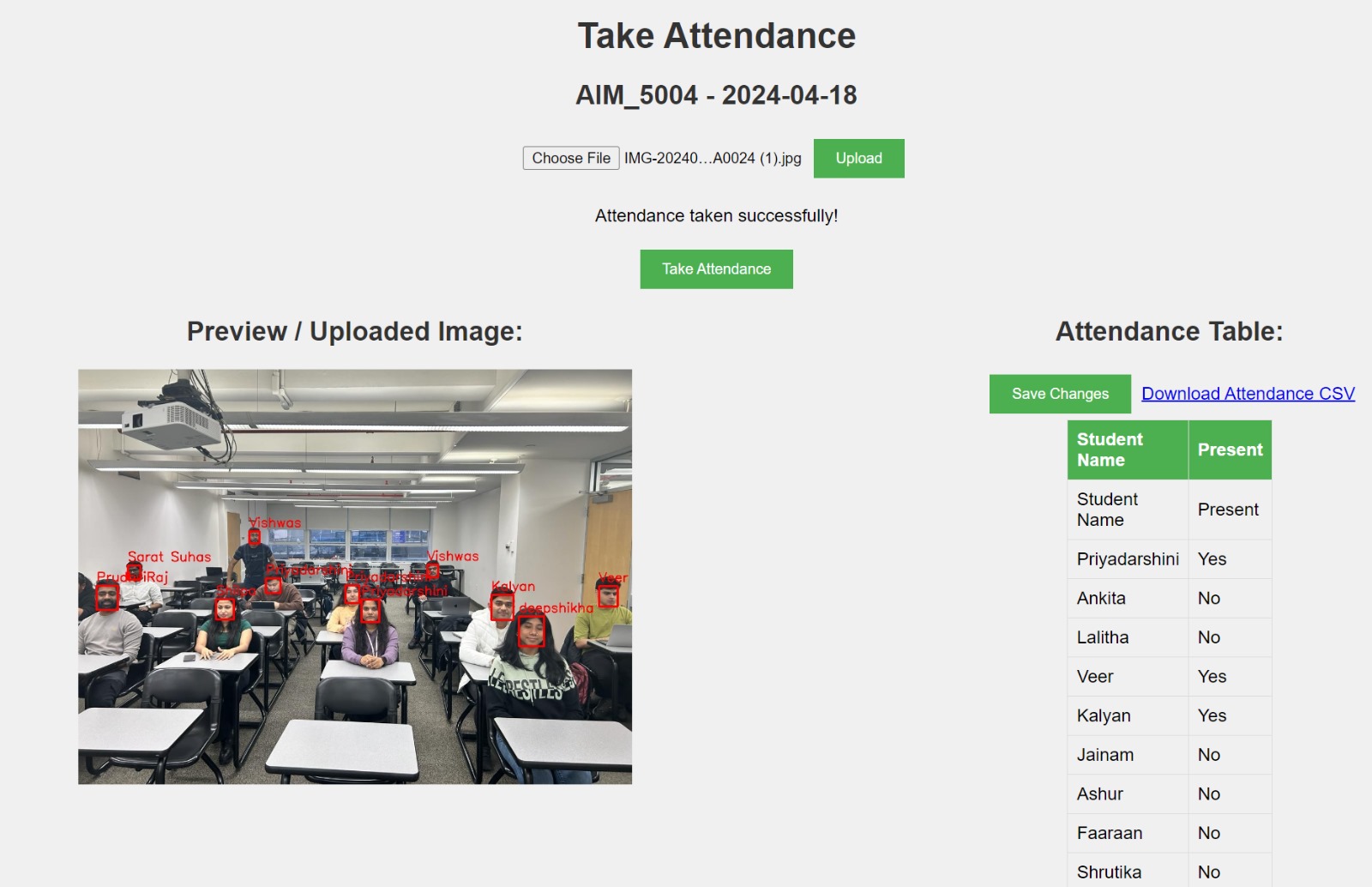}
  \caption{Screen of modifying, saving or downloading attendance CSV.}
  \label{fig:RecordAttendance}
\end{figure}

The record attendance screen (see Fig.~\ref{fig:RecordAttendance}) provides users the flexibility to modify and update the attendance for any record and save it. Once the required changes are done, the user can easily download the updated attendance records using the download attendance CSV file which follows a naming convention as CourseCode-mm-dd-yyyy.csv i.e., AIM\_5004-04-25-2024.csv .

\section{Results}

\begin{figure}[ht]
  \centering
  \includegraphics[width=\linewidth, height=5cm]{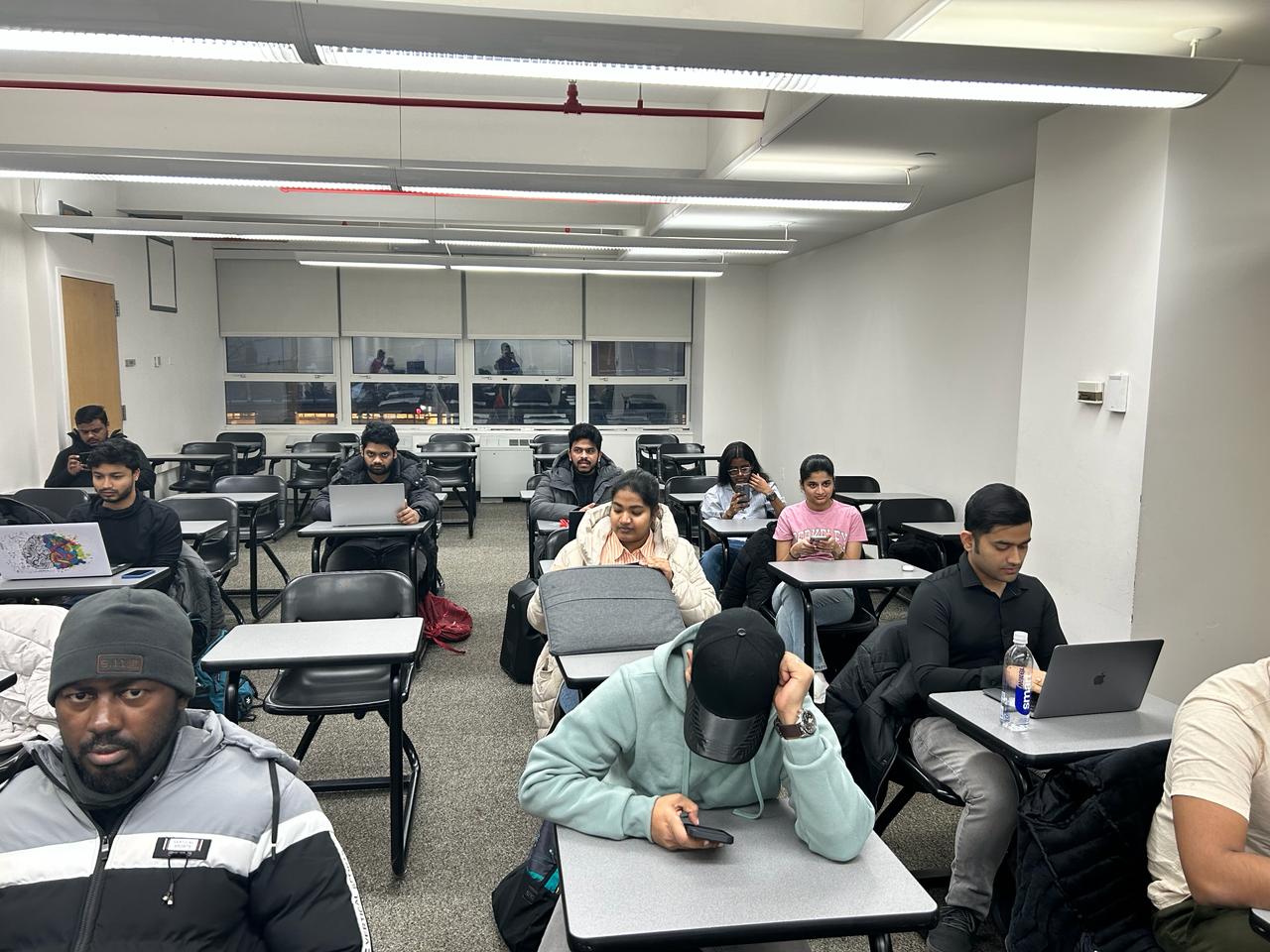}
  \caption{Ground Truth Image.}
  \label{fig:15_raw}
\end{figure}

Our experimental results comprise multiple combinations of face detection and face recognition techniques in a classroom setting with varying numbers of students, ranging from 2 to 27 students in each image. A supervised evaluation approach was followed to manually calculate the detection and recognition accuracy. For detection, each detected face was manually verified in all the test images to determine the number of faces available in each image and how many faces were undetected. Similarly, for face recognition, each image was manually checked to verify if the generated label matched the ground truth label for each student. One such sample ground truth image is provided below (see Fig.~\ref{fig:15_raw}). The overall accuracy evaluation of Facial Detection and Recognition for different models can be referenced in Table~\ref{tab:facial_results}. Additional individual image accuracy results can be found in our GitHub link. 

\subsection{Detection}
\begin{figure}[ht]
  \centering
  \includegraphics[width=\linewidth, height=5cm]{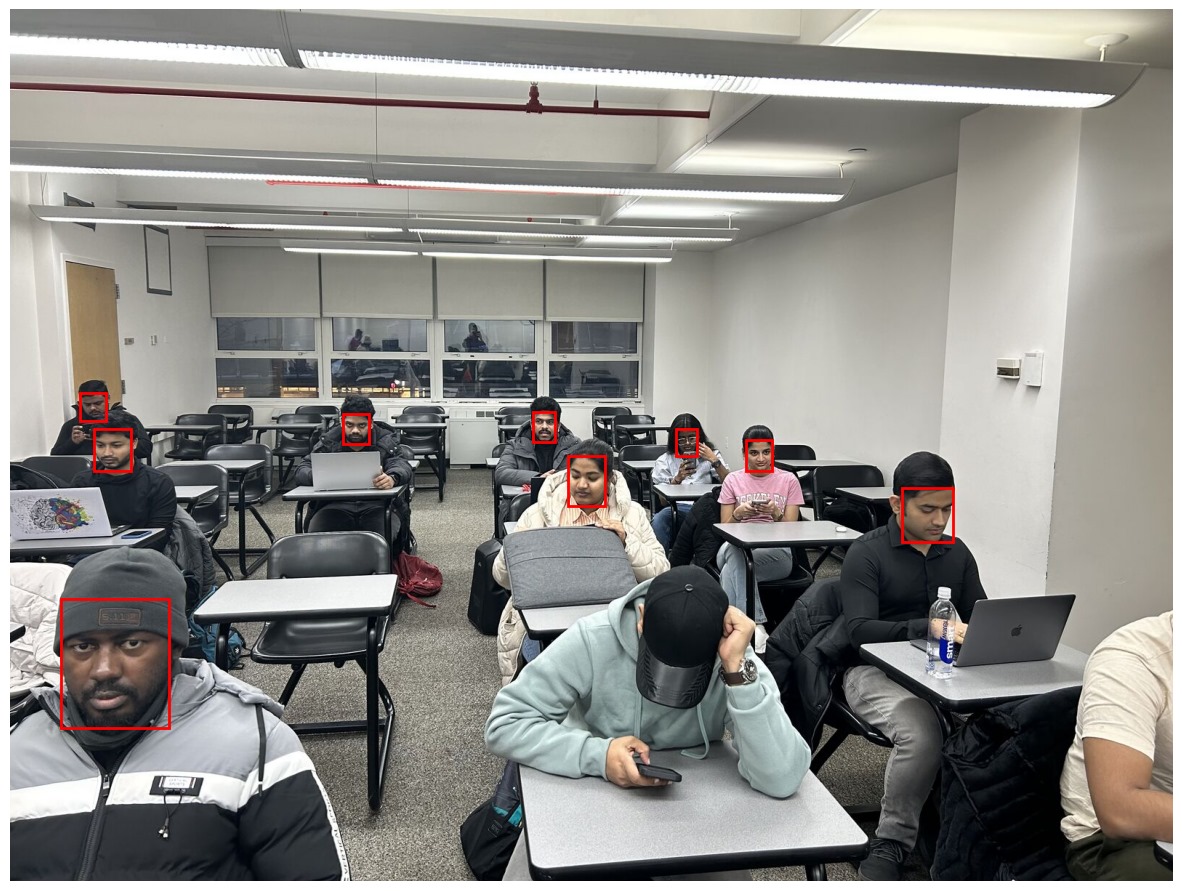}
  \caption{Retinaface Detection on group image.}
  \label{fig:15_Detection}
\end{figure}

The RetinaFace can identify nearly all the faces present in the test images (see Fig.~\ref{fig:15_Detection}). A sample image displaying the RetinaFace output on one of our collected classroom group images is shown below.

The logic to calculate detection accuracy (\%) is:
\begin{equation}
\text{Detection accuracy} = \frac{\text{Number of detected faces}}{\text{Total faces in image}}  \times 100.
\end{equation}

\subsection{Recognition}

Utilizing a face recognition library for the identification of faces, coupled with RetinaFace, provides remarkable results for group images in a classroom setting (see Fig.~\ref{fig:15_predicted}). An example demonstrating the output of face recognition is depicted below.

\begin{figure}[ht]
  \centering
  \includegraphics[width=\linewidth, height=6cm]{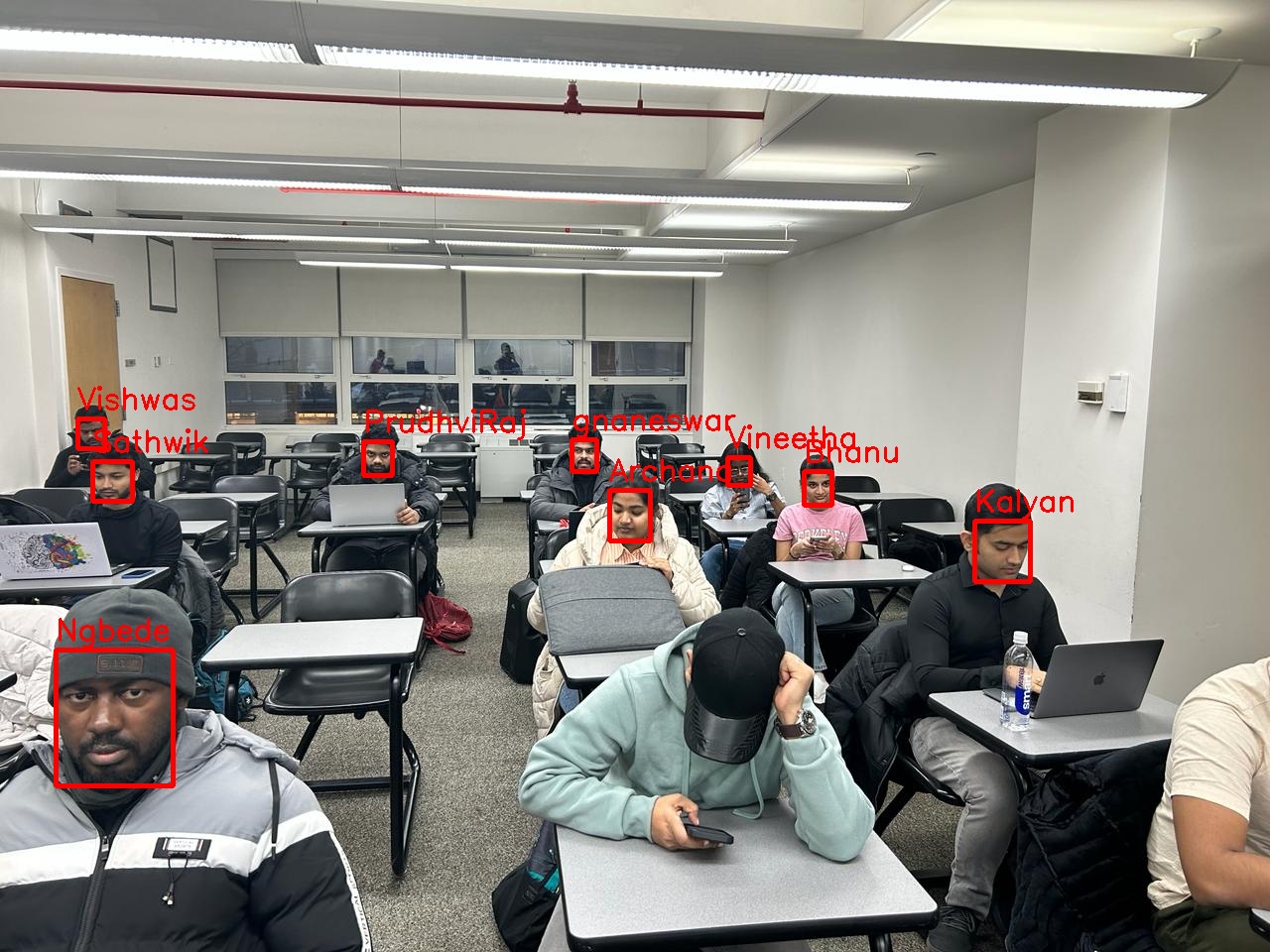}
  \caption{Face Recognition on group image.}
  \label{fig:15_predicted}
\end{figure}

The logic to calculate recognition accuracy (\%) is:

\begin{equation}
\text{Recognition accuracy} = \frac{\text{Correctly identified faces(class)}}{\text{Available faces in Class}}  \times 100.
\end{equation}

The ``available faces" refer to the actual number of students present and detectable in the classroom during attendance, not all students enrolled in the dataset. The recognition accuracy is calculated based on the system's performance in real conditions, reflecting true classroom attendance rather than theoretical maximums.

\begin{table*}[htbp]
\centering
\caption{Overall Accuracy Evaluation of Facial Detection and Recognition}
\begin{adjustbox}{max width=\textwidth}
\begin{tabular}{@{}lllclccc@{}}
\toprule
\multicolumn{1}{p{2.2cm}}{\textbf{Detection}} & \multicolumn{1}{p{2.5cm}}{\textbf{Recognition}} & \multicolumn{1}{p{2cm}}{\textbf{Image Size}} & \textbf{Cropped Embeddings} & \textbf{Detection Accuracy (\%)} & \textbf{Recognition Accuracy (\%)} & \multicolumn{1}{p{2cm}}{\textbf{Group Image Recognition}} & \multicolumn{1}{p{2cm}}{\textbf{Recognition Format - Images}} \\ \midrule
RetinaFace~\cite{retinaface} & Face Recognition~\cite{fr} & 50x50 & \ding{55} & 99.4 & 88.3 & \ding{51} & \ding{51} \\
RetinaFace~\cite{retinaface} & Face Recognition~\cite{fr} & 112x112 & \ding{55} & 99.2 & 85.3 & \ding{51} & \ding{51} \\
RetinaFace~\cite{retinaface} & Face Recognition~\cite{fr} & 600x600 & \ding{55} & 98.6 & 85.1 & \ding{51} & \ding{51} \\
Face Recognition~\cite{fr} & Face Recognition~\cite{fr} & 50x50 & \ding{51} & 65.5 & 72.3 & \ding{51} & \ding{51} \\
Face Recognition~\cite{fr} & Face Recognition~\cite{fr} & 112x112 & \ding{55} & 65.5 & 70.5 & \ding{51} & \ding{51} \\
Yolo9~\cite{yolov9_1} & Face Recognition~\cite{fr} & 112x112 & \ding{55} & 51.7 & 58.4 & \ding{51} & \ding{51} \\
RetinaFace~\cite{retinaface} & haarcascade\_frontalface~\cite{c12} & 50x50 & \ding{55} & 99.4 & 45.2 & \ding{51} & \ding{51} \\
InsightFace~\cite{deng2018arcface} & InsightFace~\cite{deng2018arcface} & 1200x1600 & \ding{55} & 99.0 & 89.2 & \ding{51} & \ding{51} \\
haarcascade\_frontalface~\cite{fr} & Face Recognition~\cite{fr} & 112x112 & \ding{55} & 91.0 & 86.0 & \ding{51} & \ding{51} \\
InsightFace~\cite{deng2018arcface} & InsightFace~\cite{deng2018arcface} & 1200x1600 & \ding{55} & 99.0 & 88.0 & \ding{51} & \ding{51} \\
Haar cascade~\cite{c12} + DoG filtering & LBPH~\cite{c12} & - & \ding{55} & 98.3 & 87.0 & \ding{55} & \ding{55} \\ 
RetinaFace[\textbf{Ours}]~\cite{retinaface} & Face Recognition~\cite{fr} & \textbf{50x50} & \textbf{\ding{51}} & \textbf{99.4} & \textbf{90.3} & \textbf{\ding{51}} & \textbf{\ding{51}} \\ \bottomrule
\end{tabular}
\label{tab:facial_results}
\end{adjustbox}
\end{table*}

Below are the key terms and their definitions used in the context of our analysis for individual test images. 

\noindent\textbf{Total Faces:} Number of total faces present in an image.\newline
\textbf{Detected Faces:} Number of Correctly detected faces in an image by RetinaFace.\newline
\textbf{Available in Training:} Number of students in the image that are also part of the training dataset.\newline
\textbf{Correctly Identified:} Number of students correctly identified by face recognition which is also part of the training dataset.\newline
\textbf{Accuracy:} Number of correctly identified faces out of number of available faces in training.

\section{Implementation details}

All the tasks for this research are conducted using the Google Colab environment, leveraging a folder structure to keep the files separate from one another. We created four folders: training, testing, processing, and prediction. The first folder, `training', contains all the images of the students in their respective folders collected during the data collection stage. The `testing' folder contains all the classroom images with students, which will be used to carry out predictions. The `processed' folder contains the images that were processed after completion of the Region of Interest (ROI) step. Finally, the `prediction' folder will contain the final images with bounding boxes and corresponding identified labels. The training and testing of the model are conducted within the Google Colab environment, leveraging the computational power of a T4 GPU. This setup offers significant advantages, including access to powerful GPU resources for accelerating model training and inference tasks.

\section{Discussions}

Our proposed framework, ClassVision, achieved remarkable results in performing facial detection and recognition tasks in a classroom setting. ClassVision represents a novel experimentation that involves the detection and recognition of faces in a classroom setting. According to our research, existing solutions focus on students or individuals being closely positioned in front of the camera to capture attendance. However, our research results are independent of this constraint. Manual evaluation of individual images helps us derive more insights into the reasons for incorrect predictions, aiding in further evaluation. During the evaluation of each incorrect prediction, we found certain factors such as obstruction in front of the student in the testing image (like a laptop, bag, or bottle), only half of the face being visible in the image, or a person being too far sitting at the back, making it hard to recognize them even manually; this is particularly observed in large classrooms. We acknowledge that there is a certain bias in our training dataset due to factors such as individuals wearing caps, facial hair, thick beards and limited images of left and right views of the face, which could lead to incorrect predictions. 

To address these issues, more robust methods can be used to capture training images of higher resolution. Better high-resolution cameras can be used to capture students sitting at the back with clear facial features. Issues regarding caps and glasses can be resolved by collecting training datasets with and without caps and glasses. We can further enhance the accuracy of recognition by developing new models. Our solution has achieved significant success in detection and recognition within classroom settings, paving the way for further research in this field. Future enhancements could involve integrating our system with educational software, such as Learning Management Systems (LMS), to automate administrative tasks and enrich data analytics. Additionally, we aim to extend this research to more advanced recognition algorithms, potentially incorporating multimodal biometric systems like voice or gesture recognition, to enhance accuracy and adaptability across diverse classroom environments.

\section{Conclusion}
We present a novel approach, ClassVision, that leverages specific preprocessing steps to prepare training images, which will be used to create effective embeddings of known images using RetinaFace. The key component of our framework is the extraction of the Region of Interest (ROI) from training images to remove unused facial features and consider only the facial features. Further resizing of the extracted Region of Interest to 50x50 dimensions helps to facilitate better comparison with the extracted images during the facial recognition step. Extensive experiments demonstrate that our proposed framework is capable of detecting and correctly recognizing faces in a classroom setting without the need for individuals to be present in front of the camera, and it achieves state-of-the-art performance. 


\small
\bibliographystyle{unsrt}
\bibliography{reference}

\end{document}